\documentclass[aps,prl,reprint,groupedaddress,showpacs,amsmath,amssymb]{revtex4-1}
\usepackage{bm}
\usepackage{graphicx}

\usepackage{color}

\newcommand{\sh}{{}} 
\newcommand*\braket[1]{\langle{#1}\rangle}
\newcommand*\sub[1]{_{\textnormal{#1}}}

\newcommand*\pdiff[3][{}]{\frac{\partial^{#1}{#2}}{\partial {#3}^{#1}}}

\DeclareMathOperator{\Tr}{Tr}
\renewcommand\Re{\operatorname{Re}}
\renewcommand\Im{\operatorname{Im}}

\begin{document}

\title{
Universal Property of Quantum Measurements of Equilibrium Fluctuations\\ 
and Violation of Fluctuation-Dissipation Theorem}

\author{Kyota Fujikura}
\email{fujikura@ASone.c.u-tokyo.ac.jp}
\author{Akira Shimizu}
\email{shmz@ASone.c.u-tokyo.ac.jp (contact author)}
\affiliation{Department of Basic Science, University of Tokyo, 3-8-1 Komaba, Meguro, Tokyo 153-8902, Japan}

\date{\today}

\begin{abstract}
For macroscopic quantum systems,
we study what are measured 
when equilibrium fluctuations of macrovariables are measured
{\sh in an ideal way that mimics classical ideal measurements
as closely as possible.
We find that the symmetrized time correlation (symTC) is always obtained 
for such measurements.
As an important consequence, 
we show that 
the fluctuation-dissipation theorem (FDT) 
is partially 
violated as a relation between observed quantities
in macroscopic quantum systems
even if measurements are made in such an ideal way.}
\end{abstract}






\maketitle

When temporal fluctuations of observables are measured in quantum systems, 
disturbances by the measurement play crucial roles
\cite{glauber,gardiner,MW,KSreview,wiseman2010,NC}.
Suppose that one measures fluctuation of $\hat{A}$
by measuring $\hat{A}$ at time $t=0$ and subsequently at time $t>0$.
Then the disturbance by the first measurement affects the outcome of the 
second one.
Consequently, fluctuation of $\hat{A}$, expressed by its time 
correlation, depends strongly on the way of measuring it.
For example, 
the normal-order time correlation and antinormal-order time correlation are obtained, respectively, 
when a photon field is measured with a photodiode and 
a quantum counter \cite{glauber,gardiner,MW}.
This means, for example, that 
the former cannot measure the zero-point fluctuation \cite{MW}.

Such strong dependence on measuring apparatuses should also be present 
in measurements of equilibrium fluctuations
in macroscopic quantum systems.
This may affect fundamental relations such as 
the FDT, 
{\sh 
which is a universal relation between 
response functions (that characterize responses to driving forces)
to 
equilibrium fluctuations (that are expressed by time correlations)}
\cite{Jhonson,Nyquist,CW1951,Kubo,Kubo2,KTH}.
%
%
However, the disturbances by measurements were 
completely disregarded when deriving the FDT \cite{Nyquist,CW1951,Kubo,Kubo2,KTH}. 
{\sh It is obvious that violent measurements of time correlations 
will lead to apparent violation of the FDT because of strong
disturbances.
Hence, one would expect that the FDT would hold only when 
certain `ideal' measurements are made for time correlations.}
For classical systems, ideal measurements are 
those which do not disturb the system at all.
For quantum systems, however, 
such measurements are generally impossible.
Hence, 
`ideal' measurements are those which mimic 
classical ideal measurements as closely as possible.
It is expected that 
the FDT should hold 
{\sh when such `ideal' measurements are made.}
Is this expectation true?

A similar problem exists in 
Onsager's regression hypothesis \cite{Onsager1,Onsager2}.
While Refs.~\cite{Nakajima56,KYN} claimed 
its consistency with the Kubo formula \cite{Kubo}, 
other works claimed inconsistency \cite{KY,Talkner,Ford}.
To settle this controversy in an operational manner, 
disturbances by measurements should be considered,
which were disregarded in these works.

%
%

In this paper, 
we study,
using modern theory of quantum measurement 
\cite{glauber,gardiner,MW,KSreview,wiseman2010,NC},
what is observed when 
{\sh `ideal' measurements are made on equilibrium fluctuations.}
We find the universal answer that 
the symTC is always observed. 
As a striking consequence, 
the FDT and the regression hypothesis 
are violated as relations between observed quantities.
Furthermore, 
we show that in the post-measurement states, 
{\sh unlike in the Gibbs states,}
expectation values of macroscopic observables evolve with time.
Such states should be realized in experiments on equilibrium fluctuations.

{\em Assumptions on equilibrium state.---}
We consider a $d$-dimensional macroscopic quantum system 
($d \geq 1$), 
whose size is characterized by $N$ (such as the number of spins).
The pre-measurement state,
represented by the canonical Gibbs state $\hat{\rho}\sub{eq}$, 
is its equilibrium state
at finite temperature $T$ ($=1/\beta$).
%
%

We assume that the correlation between any local observables 
at two points $\bm{r}$ and $\bm{r}'$ decays 
faster than $1/|\bm{r} - \bm{r}'|^{d+\epsilon}$ ($\epsilon>0$) 
with increasing $|\bm{r} - \bm{r}'|$.
This assumption is believed to hold generally except in special regions,
 such as critical points, in the thermodynamic configuration space.
Under this assumption, 
the static equilibrium fluctuation, 
$\delta A\sub{eq} 
= (\braket{ (\Delta \hat{A})^2  }\sub{eq})^{1/2}$,
of any additive 
observable $\hat{A}$ is $O(\sqrt{N})$ \cite{LL,SM2002,SM2005}.
Hereafter 
$\braket{\ \cdot\ }\sub{eq} \equiv \Tr[ \hat{\rho}\sub{eq} \ \cdot\ ]$
and 
$\Delta$ denotes 
the shift from the equilibrium value, i.e., 
$\Delta \hat{A} \equiv \hat{A} - \braket{ \hat{A} }\sub{eq}$.
Furthermore, 
with some reasonable additional assumptions (see Ref.~\cite{SM}), 
the `quantum central limit theorem' \cite{SM,goderis1989mixing,matsui2002,jaksic2008,Mthesis} holds,
which enables us to derive the following universal results.
Its statement, 
discussion on related theorems \cite{Lieb,Nachtergaele,LRB_cont},
and details of the calculations
are described in Ref.~\cite{SM}.

%
%
%

{\em Quasiclassical measurement.---}
{\sh As discussed above, 
certain `ideal' measurements, which mimic classical ones, should be made
to get time correlations correctly.
We call such measurements {\em quasiclassical}.}
To be definite, we define them 
as minimally-disturbing, homogeneous, and unbiased 
quantum measurements with moderate magnitudes of errors,
as follows.

To observe fluctuation of an additive observable $\hat{A}$,
its measurement error $\delta A\sub{err}$ should be smaller 
than $\delta A\sub{eq}$.
On the other hand, $\delta A\sub{err}$ should not be too small 
because otherwise disturbances by the measurement would be too large
so that the measurement would be much different from classical ideal ones.
Therefore, we consider measurements in which 
$\delta A\sub{err} = \varepsilon \delta A\sub{eq}$, 
where $\varepsilon$ is independent of $N$. 
Although we are interested in the case where $\varepsilon < 1$, 
the following results hold also for larger $\varepsilon$
{\sh (which occurs, e.g., when interaction with 
the measuring apparatus is weak \cite{KSreview}).}
Considering $\delta A\sub{eq} = O(\sqrt{N})$ as mentioned above, 
we require more concretely 
(i) the measurement operator 
{\sh (that gives the state just after measurement \cite{wiseman2010,NC})}
scales as $\sqrt{N}$ with increasing $N$. 
A quasiclassical measurement should also possess 
the following reasonable properties: 
(ii) Minimally disturbing, i.e, 
the only disturbance of the system is the necessary backaction 
determined by the probability operator \cite{wiseman2010}.
(iii) Homogeneous, i.e., 
the measurement operator is a function of 
$\hat{A}-A_\bullet$,
where $A_\bullet$ is the outcome of the measurement.
{\sh This implies, e.g., a reasonable condition that 
$\delta A\sub{err}$ is independent of $A_\bullet$.}
(iv) Unbiased, i.e, $\overline{A_\bullet} = \braket{\hat{A}}\sub{eq}$,
where $\overline{\cdots}$ denotes average over many runs of experiments.
From (i)-(iv), 
the measurement operator should take the form $f( \hat{a} - a_\bullet )$
in terms of scaled quantities 
$\hat{a} = \hat{A}/\sqrt{N}$
and 
$a_\bullet = A_\bullet/\sqrt{N}$.
Here, $f(x)$ is a real function independent of $N$ 
such that 
$\int |f(x)|^2 dx = 1, \int x |f(x)|^2 dx = 0$.
Hence, 
the probability density for getting $a_\bullet$ is
\begin{equation}
p(a_\bullet)
=
\big\langle \big\{ 
f(\hat{a} - a_\bullet) \big\}^2 \big\rangle\sub{eq},
\label{eq:prob_A}
\end{equation}
and the post-measurement state for the outcome $a_\bullet$ is \cite{rmk:f}
\begin{equation}
\hat{\rho}(a_\bullet) 
=
f( \hat{a} - a_\bullet ) 
\hat{\rho}\sub{eq}
f( \hat{a} - a_\bullet )/p(a_\bullet).
\label{eq:post_state}
\end{equation}
It seems reasonable to require also 
(v) $f(x)$ behaves well enough,
e.g., it vanishes quickly with increasing $|x|$
(such as the Schwartz functions \cite{terziogglu1969,homander1990,reed1980,stein2003}; 
see Ref.~\cite{SM} for details).
Although (ii) requires also that 
$f$ should be nonnegative up to an irrelevant phase factor \cite{wiseman2010}, 
all the following results hold without this additional condition.
Then, the measurement error $\delta a\sub{err}$ ($=\delta A\sub{err}/\sqrt{N}$)
is given by 
$ 
(\delta a\sub{err})^2 = \int x^2 |f(x)|^2 dx = O(1),
$ 
in consistency with $\delta A\sub{err} = \varepsilon \delta A\sub{eq}$.
A typical $f$ is gaussian \cite{gaussian_case},
$ 
f(x) = (2\pi w^2)^{-1/4} \exp (- x^2/4w^2)
$ 
with $w=O(1)>0$,
for which $\delta a\sub{err} = w$.

{\em Which time correlation is measured?---}
Although all the following results 
hold in the thermodynamic limit, 
we do not write $\lim_{N \to \infty}$ explicitly.
Suppose that the macroscopic system was in 
a Gibbs state $\hat{\rho}\sub{eq}$ for $t<0$,
and that an additive observable $\hat{A}$ is measured at $t=0$.
The outcome 
$a_\bullet$ ($= A_\bullet/\sqrt{N}$)
distributes according to Eq.~(\ref{eq:prob_A}), for which we have
\begin{equation}
p(a_\bullet) 
= 
\int 
{
\left|f(x)\right|^2
\over (2\pi \delta a\sub{eq}^2)^{1/2}
}
\exp\left[
- 
{
(x+\Delta a_\bullet)^2
\over 
2\delta a\sub{eq}^2
}
\right]
dx,
\label{eq:prob_A_by_qclt}
\end{equation}
where 
$\delta a\sub{eq}^2 \equiv (\delta A\sub{eq})^2/N$
and $\Delta a_\bullet = a_\bullet -\braket{\hat{a}}\sub{eq}$.
This means that $p(a_\bullet)$ is a convolution of 
the distribution in $\hat{\rho}\sub{eq}$ and the shape $|f(x)|^2$ of 
the measurement operator.
We also see that 
the variance scales as 
$\delta a_\bullet^2 \equiv \overline{(\Delta a_\bullet)^2} = O(1)$,
in consistency with 
$\delta a\sub{eq}^2 = O(1)$ and $\delta a\sub{err}^2 = O(1)$.

The post-measurement state $\hat{\rho}(a_\bullet)$ deviates from 
 $\hat{\rho}\sub{eq}$, 
hence the expectation values,
denoted by $\braket{\ \cdot\ }_{a_\bullet}$, of observables 
evolve with time.
We investigate the time evolution of an additive observable $\hat{B}$
($=\sqrt{N} \hat{b}$) for $t > 0$.
For
$
\braket{\Delta \hat{b}(t)}_{a_\bullet}
\equiv
\braket{\hat{b}(t)}_{a_\bullet}
-
\braket{\hat{b}}\sub{eq}
$,
we find \cite{rigour,gaussian_case} 
\begin{equation}
\braket{\Delta \hat{b}(t)}_{a_\bullet}
=
-
\Theta(t)
\langle 
\mbox{$\frac{1}{2}$}
\{ \Delta \hat{a}, \Delta \hat{b}(t) \}
\rangle\sub{eq}
(\ln p)',
\label{eq:<B>}
\end{equation}
where
$\Theta(t)$ is the step function,
$
\langle
\frac{1}{2} 
\{ \hat{X}, \hat{Y}(t) \}
\rangle\sub{eq}
\equiv
\langle 
\frac{1}{2}
(\hat{X} \hat{Y}(t) + \hat{Y}(t) \hat{X})
\rangle\sub{eq}
$
the symTC, 
$\hat{b}(t) = e^{i \hat{H} t / \hbar} \hat{b} e^{-i \hat{H} t / \hbar}$
the Heisenberg operator, 
and 
$(\ln p)' = d \ln p /d a_\bullet$.
Since $p'$ vanishes at $\Delta a_\bullet =0$ 
from Eq.~(\ref{eq:prob_A_by_qclt}), 
the rhs of Eq.~(\ref{eq:<B>}) 
is linear in $\Delta a_\bullet$ for small $|\Delta a_\bullet|$.

Averaging Eq.~(\ref{eq:<B>}) over $a_\bullet$,
we find \cite{rigour}
\begin{equation}
\overline{\braket{\Delta \hat{b}(t)}_{a_\bullet}} = 0,
\mbox{ i.e., }
\overline{\braket{\hat{b}(t)}_{a_\bullet}} = \braket{\hat{b}}\sub{eq}
\label{eq:<B>av}
\end{equation}
for all additive observable $\hat{B} = \sqrt{N} \hat{b}$
at all $t > 0$.
That is, 
the measurement does not cause any systematic 
disturbance on $\hat{b}$,
in consistency with our requirements on quasiclassical measurements.

With increasing $t$, 
$\braket{\hat{b}(t)}_{a_\bullet}$
relaxes to 
$\braket{\hat{b}}\sub{eq}$
if the system possesses 
`mixing property' in the sense that 
\begin{equation}
\lim_{t \to \infty}
\langle 
\mbox{$\frac{1}{2}$}
\{ \Delta \hat{a}, \Delta \hat{b}(t) \}
\rangle\sub{eq}
= 0,
\label{eq:mixingsymTC}\end{equation}
where $\lim_{N \to \infty}$ is taken before $\lim_{t \to \infty}$.
After the relaxation, 
one cannot distinguish 
$\hat{\rho}(a_\bullet)$ from $\hat{\rho}\sub{eq}$
by measuring any additive observables.
Equation (\ref{eq:<B>}) shows that 
the relaxation process after the measurement is governed 
by the symTC.
To see it clearly, 
we calculate the correlation between 
$a_\bullet$ and $\braket{\hat{b}(t)}_{a_\bullet}$.
Multiplying Eq.~(\ref{eq:<B>}) with $\Delta a_\bullet$, 
and averaging over $a_\bullet$, 
we find \cite{rigour}
\begin{equation}
\overline{
\Delta a_\bullet
\braket{\Delta \hat{b}(t)}_{a_\bullet}
}
=
\Theta(t)
\langle 
\mbox{$\frac{1}{2}$}
\{ \Delta \hat{a}, \Delta \hat{b}(t) \}
\rangle\sub{eq}.
\label{eq:corr.ab}
\end{equation}
This is a universal result 
independent of choice of $f(x)$, 
and tells us the operational meaning:
{\em When one measures a time correlation quasiclassically,  
what he observes
is the symTC} rather than many other time correlations 
which reduce to the same classical correlation as $\hbar \to 0$.

This result might look contradictory to 
some experiments \cite{glauber,gardiner,MW,ec1,ec2,ec3,ec4,ec5,ec6}.
However, those measurements are not quasiclassical because 
{\sh they destruct the states by absorbing quanta.}
%
%
If, e.g., 
heterodyning techniques \cite{Koch1982,Milburn} or
quantum non-demolition 
photodetectors 
\cite{PSJmeeting,IQEC,QNDpra,QND1,QND2} are used, 
the symTC will be observed.

{\em Violation of FDT.---}
%
The above finding has a great impact on 
nonequilibrium physics.
The Kubo formula \cite{Kubo} gives 
the response function, 
which describes response of a quantum system to an external force,
by the canonical time correlation, 
denoted by $\langle \, \cdot \, ; \, \cdot \, \rangle\sub{eq}$,  
of certain additive observables as \cite{SM}
\begin{equation}
\Phi_{ba}(t) = 
\Theta(t)
\beta \langle 
\Delta \hat{a}; \Delta \hat{b}(t)
\rangle\sub{eq}.
\label{eq:LRT_Phi}
\end{equation}
Here, the step function $\Theta(t)$ comes from the causality.
This formula has been regarded 
as the FDT for quantum systems \cite{Kubo,Kubo2,KTH}.
In its derivation, however, 
disturbances by quantum measurements were neglected, 
although they should be considered seriously \cite{Takahashi}.
Therefore, we here take Eq.~(\ref{eq:LRT_Phi}) just 
as a recipe to obtain the response function,
as discussed in Ref.~\cite{SM}, 
while measured fluctuation may possibly be described 
by a different expression \cite{SK}.
%
%
Then a question arises; 
does the FDT hold as relations between measured quantities
in quantum systems? 

To answer this fundamental question,
%
we calculate the Fourier transform of $\Phi_{ba}(t)$, 
which is the admittance denoted by $\chi_{ba}(\omega)$, 
and that of the measured time correlation Eq.~(\ref{eq:corr.ab}), 
denoted by $S_{ba}(\omega)$
(whereas the Fourier transform of the symTC without $\Theta(t)$ is 
denoted by $\tilde{S}_{ba}(\omega)$).
We express the results in terms 
of the symmetric and antisymmetric parts, 
denoted by $+$ and $-$, respectively, 
such as $\chi_{ba}^\pm \equiv (\chi_{ba} \pm \chi_{ab})/2$.
For the real parts, we have \cite{similar}
\begin{align}
\Re \chi_{ba}^+(\omega)
&= 
\beta \Re S_{ba}^+(\omega)/I_\beta(\omega),
\label{eq:chi+}
\\
\Re \chi_{ba}^-(\omega)
&=
\beta \Re S_{ba}^-(\omega)
\nonumber\\
+
\beta \int_{-\infty}^{\infty} 
& {\mathcal{P} \over \omega' - \omega}
\left[ 1-  {1 \over I_\beta(\omega')} \right]
i \tilde{S}_{ba}^-(\omega')
{d\omega' \over 2 \pi}.
\label{eq:chi-}
\end{align}
Here, $\mathcal{P}$ denotes the principal value,
and
$I_\beta(\omega) \equiv (\beta \hbar \omega / 2) \coth(\beta \hbar \omega / 2)$.
[For the imaginary parts, replace
$\Re \chi_{ba}^\pm$, $\Re S_{ba}^\pm$ and $i \tilde{S}_{ba}^-$
with 
$\Im \chi_{ba}^\mp$, $\Im S_{ba}^\mp$ and $\tilde{S}_{ba}^+$,
respectively.]
Since $I_\beta(\omega) \neq 1$ at finite $\omega$,
we find that 
{\em 
the FDT is violated at finite $\omega$, 
when comparing the observed equilibrium fluctuation
and the observed admittance} 
{\sh even if the measurement is quasiclassical (i.e., even if it 
mimics classical ideal measurement.)}


One might expect that the FDT would recover 
in the `classical regime' where $\hbar \omega \ll k\sub{B} T$, 
because $I_\beta(\omega) \to 1$.
We examine this expectation by 
studying the case of $\omega=0$
(for which 
$\chi_{ba}^\pm(0) = \Re \chi_{ba}^\pm(0)$ 
because $\Phi_{ba}(t)$ is real).
From Eqs.~(\ref{eq:chi+}) and (\ref{eq:chi-}), 
we find that even at $\omega=0$
the FDT is recovered only for the symmetric part.
For the antisymmetric part, 
the causality in Eq.~(\ref{eq:LRT_Phi})
(represented by $\Theta(t)$)
convolutes different 
frequencies as
\begin{equation}
{\chi_{ba}^-(0) \over \beta} - S_{ba}^-(0)
=
\int_{-\infty}^{\infty} 
{\mathcal{P} \over \omega}
\left[ 1-  {1 \over I_\beta(\omega)} \right]
i \tilde{S}_{ba}^-(\omega)
{d\omega \over 2 \pi}.
\label{eq:chi-S}
\end{equation}
Since $\tilde{S}_{ba}^-(\omega)$ is a pure-imaginary odd function \cite{KTH}, 
the rhs of Eq.~(\ref{eq:chi-S}) does not vanish in general. 
Therefore, in general, 
{\em the FDT 
is violated even in the `classical regime' for the antisymmetric part}
{\sh even if the measurement is quasiclassical.}
Note that 
there are two ways to reach the 'classical regime' $\beta \hbar \omega \ll 1$.
One is to take $\hbar \to 0$, where the system becomes classical and the violation disappears.
The other is to take $\omega \to 0$ while keeping $\hbar$ constant, where
the violation occurs.
Therefore, the violation is a genuine quantum effect.

{\em Experiments.---}
As an example, we compare the conductivity tensor \cite{SM}, 
$
\sigma_{\mu \nu} (\omega) 
= 
\beta \int_0^\infty 
\langle 
\hat{j}_\nu; \hat{j}_\mu(t)
\rangle\sub{eq}
e^{i \omega t} dt
$,
with the Fourier transform of the measured equilibrium fluctuation, 
$
S_{\mu \nu}(\omega)
\equiv
\int_0^\infty 
\langle 
\mbox{$\frac{1}{2}$}
\{ \hat{j}_\nu, \hat{j}_\mu(t) \}
\rangle\sub{eq}
e^{i \omega t} dt
$,
where $\hat{j}_\nu$ denotes 
the $\nu$ component ($\nu=x,y,z$) of 
the total current divided by $\sqrt{N}$. 
%
%

Koch et al.\  \cite{Koch1982} measured the 
diagonal ($\mu=\nu$) elements of the equilibrium fluctuation of a circuit 
by using the heterodyning technique \cite{Milburn}.
{\sh Their measurement is closer to quasiclassical
than those of Refs.~\cite{ec1,ec2,ec3,ec4,ec5,ec6}
because it does not destruct states by absorbing quanta, 
as discussed above.}
%
%
Hence, it can be regarded as a pioneering work
about the equilibrium fluctuation obtained by quasiclassical measurements.
Its results are consistent with Eq.~(\ref{eq:chi+}).

For the off-diagonal ($\mu \neq \nu$) elements,
no experiments have been reported 
to the authors' knowledge.
We propose to examine the $\omega=0$ case, 
Eq.~(\ref{eq:chi-S}), 
by studying 
$\sigma_{x y}(0)$ in the presence of 
a magnetic field parallel to the $z$ axis.
When the system is invariant under rotation by $\pi/2$ about the $z$ axis, 
the obvious symmetry $\sigma_{x y} = - \sigma_{y x}$
and Eq.~(\ref{eq:chi-S}) yield
\begin{equation}
{\sigma_{x y} (0) \over \beta} - S_{x y}(0)
=
\! \!
\int_{-\infty}^{\infty} 
{\mathcal{P} \over \omega}
\left[ 1-  {1 \over I_\beta(\omega)} \right]
i \tilde{S}_{x y}^-(\omega)
{d\omega \over 2 \pi}.
\label{eq:sigma-S}
\end{equation}
Therefore, even at $\omega=0$ the FDT 
is violated 
for the off-diagonal (Hall) conductivity.
This violation will be confirmed 
by measuring
independently
$\sigma_{xy}(0)$ and the correlation 
$\overline{
j_{y \bullet}
\lower.3ex\hbox{$\braket{\hat{j}_x(t)}_{j_{y \bullet}}$}
}
$,
which gives $S_{xy}(\omega)$ according to Eq.~(\ref{eq:corr.ab}).
One can also obtain the rhs of Eq.~(\ref{eq:sigma-S})
by measuring 
$\overline{j_{\nu \bullet}^0 j_{\mu \bullet}^k}$
and
$\overline{j_{\mu \bullet}^0 j_{\nu \bullet}^k}$,
which together give $\tilde{S}_{\mu \nu}(\omega)$ 
according to Eq.~(\ref{eq:corr.a0ak}) below.

{\em Regression hypothesis.---}
References \cite{KY,Talkner,Ford} pointed out 
that Onsager's regression hypothesis \cite{Onsager1,Onsager2}
would contradict with Eq.~(\ref{eq:LRT_Phi}),
assuming 
the symTC for the time correlation of the hypothesis.
Nakajima showed that this contradiction can be removed 
if a local equilibrium state is assumed for 
the state during fluctuation \cite{Nakajima56},
and thereby derived response to non-mechanical forces \cite{KYN}.
These contradictory claims were derived from different assumptions, 
{\sh but none of them was justified satisfactorily.}
Since we have proved that 
the symTC 
is always measured in quasiclassical experiments,
the hypothesis 
{\sh cannot be valid}
as a relation between measured quantities 
in quantum systems
{\sh even if the measurements mimic classical ideal measurements.}


{\em Squeezed equilibrium state.---}
We have 
shown how quantum effects appear 
in the expectation value, 
Eqs.~(\ref{eq:<B>}) and (\ref{eq:<B>av}), 
and in the correlation, Eq.~(\ref{eq:corr.ab}).
%
%
Quantum effects appear more manifestly in 
the variance of $\hat{b}(t)$, 
which is calculated as \cite{gaussian_case} 
\begin{eqnarray}
&&
\langle (\hat{b}(t) - \langle \hat{b}(t) \rangle_{a_\bullet})^2 \rangle_{a_\bullet}
-
\delta b\sub{eq}^2
\nonumber\\
&&  = \!
\braket{ \mbox{$\frac{1}{2}$} \{ \Delta \hat{a}, \Delta \hat{b}(t) \}}\sub{eq}^2 
(\ln p)''
\! -
\braket{\mbox{$\frac{1}{2i}$} [ \hat{a}, \hat{b}(t) ]}\sub{eq}^2  
\! \! \left. \pdiff[2]{ \ln q}{y} \right|_{y=0} \! \! ,
\ \ \ \
\label{eq:dB^2}\end{eqnarray}
where 
%
%
\begin{eqnarray}
&&
q(a_\bullet,y)
\equiv
\big\langle f(\hat{a} - a_\bullet + y) f(\hat{a} - a_\bullet - y) \big\rangle\sub{eq}
\nonumber\\
&& \quad
=
\int 
{
f(x+y) f(x-y)
\over 
(2\pi \delta a\sub{eq}^2)^{1/2}
}
\exp\left[
- 
{
(x+\Delta a_\bullet)^2
\over 
2\delta a\sub{eq}^2
}
\right]
dx.
\ \
\label{eq:prob_ay}
\end{eqnarray}
We find that the relaxation process is governed by both
the symTC and the commutator time correlation
$\braket{\mbox{$\frac{1}{2i}$} [ \hat{a}, \hat{b}(t) ]}\sub{eq}$. 
If, in addition to Eq.~(\ref{eq:mixingsymTC}), the system also has the mixing property in the sense that 
$ 
\lim_{t \to \infty}
\lim_{N \to \infty}
\braket{\mbox{$\frac{1}{2i}$} [ \hat{a}, \hat{b}(t) ]}\sub{eq}^2  
= 0,
$ 
then the variance 
relaxes to 
$\delta b\sub{eq}^2$ 
with increasing $t$. 

We can see the physical meaning of 
Eq.~(\ref{eq:dB^2})
by letting $t \to +0$.
The symTC implies that $\hat{\rho}(a_\bullet)$ is squeezed
along $\hat{b}$ 
by measuring $\hat{a}$ if $\hat{b}$ correlates with 
$\hat{a}$ in 
$\hat{\rho}\sub{eq}$.
The commutator time correlation implies that the squeezing is disturbed by 
the measurement if $[\hat{a}, \hat{b}]$ is non-negligible 
in $\hat{\rho}\sub{eq}$.
Furthermore, 
the rhs of Eq.~(\ref{eq:<B>}) is $O(1)$, 
which is the same order as $\delta a\sub{eq}$.
That is, 
$\hat{\rho}(a_\bullet)$ deviates from $\hat{\rho}\sub{eq}$
only within equilibrium fluctuations,
hence 
the system remains macroscopically in the same equilibrium state.
In this `squeezed equilibrium state,'
macrovariables evolve with time 
as Eq.~(\ref{eq:<B>}), 
unlike in the Gibbs state.
It represents the state that is observed in 
quasiclassical measurements of equilibrium fluctuations.

{\em Multi-time measurements.---}
To measure the lhs's of Eqs.~(\ref{eq:<B>}) and (\ref{eq:corr.ab}),
one must perform measurements twice in each run of the experiment, 
as described in Ref.~\cite{SM}.
%
%
When one performs more measurements subsequently in each run,
what he gets is the following.

Suppose that $K+1$ additive operators 
$\hat{A}^0, \hat{A}^1, \cdots, \hat{A}^K$
are measured subsequently 
at $t = t_0, t_1, \cdots, t_K$, 
respectively,
and the outcomes are 
$A^0_\bullet, A^1_\bullet, \cdots, A^K_\bullet$
($= \sqrt{N} a^0_\bullet, \sqrt{N} a^1_\bullet, \cdots, \sqrt{N} a^K_\bullet$).
Here, $K=O(1)$ and $0 = t_0 < t_1 < \cdots < t_K$.
Let the measurement operator for 
$\hat{A}^j = \sqrt{N}\hat{a}^j$
be $f_j( \hat{a}^j - a^j_\bullet )$,
where $f_j(x)$ satisfies the aforementioned conditions for $f(x)$.
The state at $t=t_j$ is obtained by applying 
$
f_j( \hat{a}^j \!\! - \! a^j_\bullet )
\cdots \!
e^{-i \hat{H} (t_2-t_1) / \hbar} 
f_1( \hat{a}^1 \!\! - \! a^1_\bullet ) 
e^{-i \hat{H} t_1 / \hbar} 
f_0( \hat{a}^0 \!\! - \! a^0_\bullet )
$
%
%
and its conjugate to $\hat{\rho}\sub{eq}$
from the left and right, respectively. 
We find that 
\begin{equation}
\overline{\Delta a^j_\bullet} = 0,
\mbox{ i.e., } 
\overline{a^j_\bullet} = \braket{\hat{a}^j}\sub{eq}
\ \mbox{ for all $j$}.
\label{eq:multi_time_average}
\end{equation}
For the correlations for $0 \leq j \leq k \leq K$, we get
\begin{eqnarray}
&&
\overline{\Delta a^j_\bullet \Delta a^k_\bullet}
=
\langle
\mbox{$\frac{1}{2}$}
\{ \Delta \hat{a}^j(t_j), \Delta \hat{a}^k(t_k) \}
\rangle\sub{eq}
+ \delta_{j,k} \delta a^{j \, 2}\sub{err}
\nonumber\\
&& + 
\sum_{l=0}^{j-1} F_l
\braket{\mbox{$\frac{1}{2i}$} [ \hat{a}^j(t_j), \hat{a}^l(t_l) ]}\sub{eq}
\braket{\mbox{$\frac{1}{2i}$} [ \hat{a}^l(t_l), \hat{a}^k(t_k) ]}\sub{eq},
\label{eq:calced_correlation}
\end{eqnarray}
where
$
\delta a^{j \, 2}\sub{err} = \int x^2 |f_j|^2 dx
$
and
$
F_j =  4 \int \{ f'_j(x) \}^2 dx
$ \cite{gaussian_case}.
The first term in the rhs of Eq.~\eqref{eq:calced_correlation} 
corresponds to the correlation in Eq.~(\ref{eq:corr.ab}).
The second term comes from the measurement error of each measurement,
which is absent in Eq.~(\ref{eq:corr.ab}) because 
it corresponds to the case $\delta_{j,k} = \delta_{0,1} = 0$. 
The last term represents disturbances by the measurements
that are performed before the $j$-th measurement, 
hence the summation is over $m$ such that $m < j$.
Because of this term, 
which depends on the experimental setup
represented by $\{ \hat{a}_j, f_j(x), t_j \}_j$,
the correlation deviates from the symTC
unlike the case of twice measurements, Eq.~(\ref{eq:corr.ab}).
Consequently, 
{\em the FDT is violated more drastically in this protocol of experiment}, 
e.g., 
even at $\omega=0$ both the symmetric and antisymmetric parts 
can violate the FDT.

For the special case where $j=0$ (and $k \geq 1$), 
however, the disturbance term is absent:
\begin{equation}
\overline{\Delta a^0_\bullet \Delta a^k_\bullet}
=
\langle
\mbox{$\frac{1}{2}$}
\{ \Delta \hat{a}^0, \Delta \hat{a}^k(t_k) \}
\rangle\sub{eq}
\ \mbox{ for $t_k > 0$}.
\label{eq:corr.a0ak}
\end{equation}
This coincides with Eq.~(\ref{eq:corr.ab}), although 
other measurements may be performed for $0<t<t_k$.
It is also universal,
independent of choice of $f(x)$, 
as is Eq.~(\ref{eq:corr.ab}).
Hence, in this case, the FDT is violated in the same manner  
as in the previous protocol.

In summary, 
we have studied what is measured when the equilibrium fluctuation is measured
{\sh in an ideal way that mimics 
classical ideal measurements as closely as possible,
i,e., }
such that 
disturbances are as small as possible
under the condition that the equilibrium fluctuation can 
be measured accurately enough.
{\sh We call such measurements quasiclassical.}
It is found that the symTC is obtained quite generally 
[Eqs.~(\ref{eq:corr.ab}) and (\ref{eq:corr.a0ak})].
From this finding and the causality, 
we have shown that 
the FDT is violated 
as a relation between observed quantities 
{\sh even if measurements are quasiclassical}
[Eqs.~(\ref{eq:chi+})-(\ref{eq:chi-S})].
{\sh It is violated for antisymmetric parts of response functions at all frequencies 
and for symmetric parts at high frequencies 
$\hbar \omega \gtrsim k_B T$.}
Onsager's regression hypothesis is also violated.
These striking results are genuine quantum effects.
Disturbances by measurements appear more strongly in 
the variances of macrovariables [Eq.~(\ref{eq:dB^2})]
and in the case of multi-time measurements [Eq.~(\ref{eq:calced_correlation})].
The post-measurement state 
is shown to be a time-evolving `squeezed equilibrium state,'
in which macrovariables fluctuate and relax 
[Eqs.~(\ref{eq:<B>}) and (\ref{eq:dB^2})].
It represents the state realized during 
quasiclassical measurements of equilibrium fluctuations.

{\sh 
Finally, we note that 
our results apply not only to the Gibbs states but also to 
pure states representing equilibrium states
\cite{vonN,SugitaJ,SugitaE,Popescu,Goldstein,Reimann2007,SS2012,SS2013,HSS2014}.}

\begin{acknowledgments}
We thank 
P. Reimann for informing us of Refs.~\cite{Talkner,Ford},
and R. Hatakeyama for suggesting the importance of the causality.
This work was supported by JSPS KAKENHI Nos.~26287085 and 15H05700.
\end{acknowledgments}

\clearpage

\pagestyle{empty}
\parindent=0mm
\setlength{\textwidth}{1.15\textwidth}

\begin{figure*}
\vspace{-10mm}
\hspace{-28mm}
\includegraphics[page=1,clip]{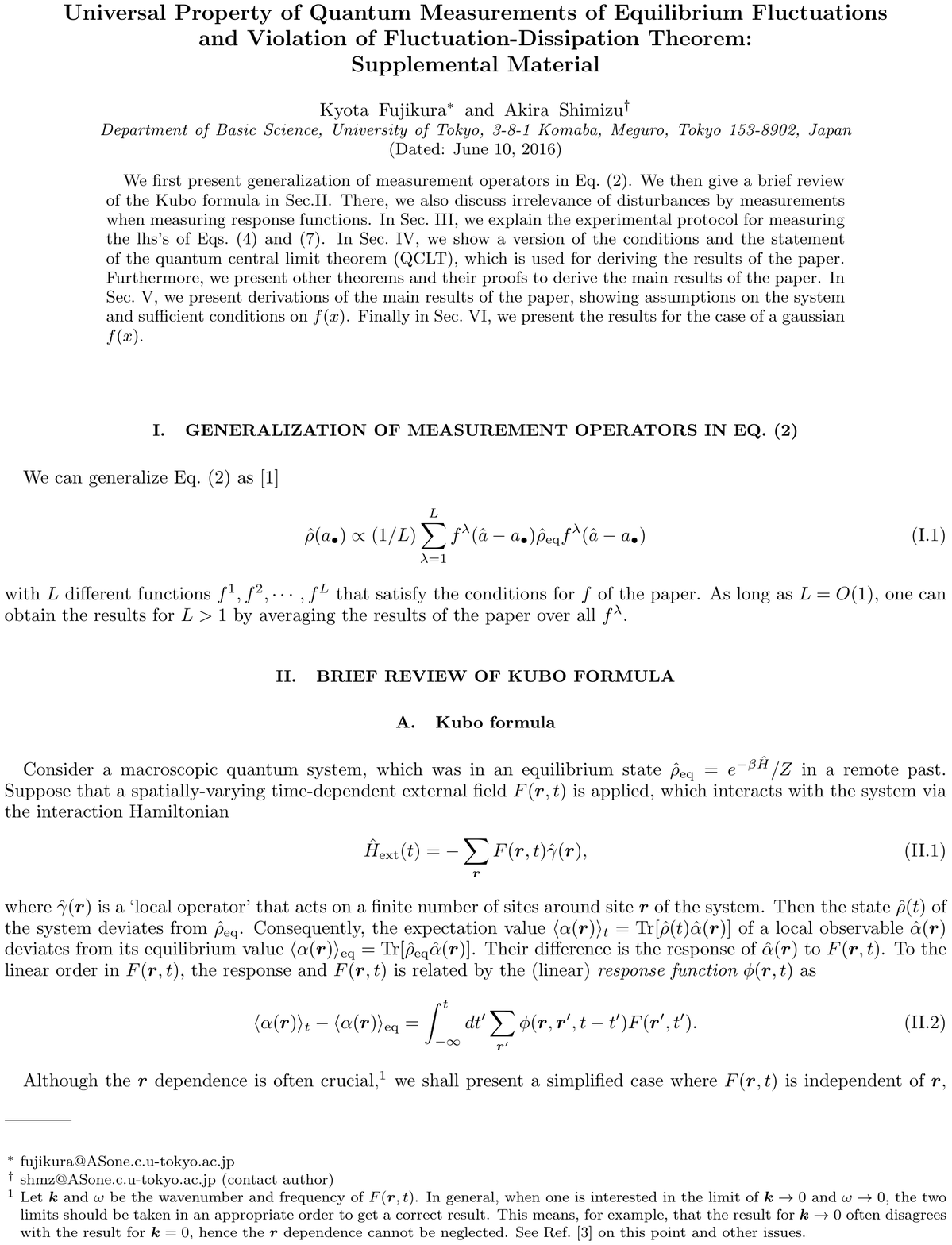}
\end{figure*}

\begin{figure*}
\vspace{-10mm}\hspace{-28mm}
\includegraphics[page=2,clip]{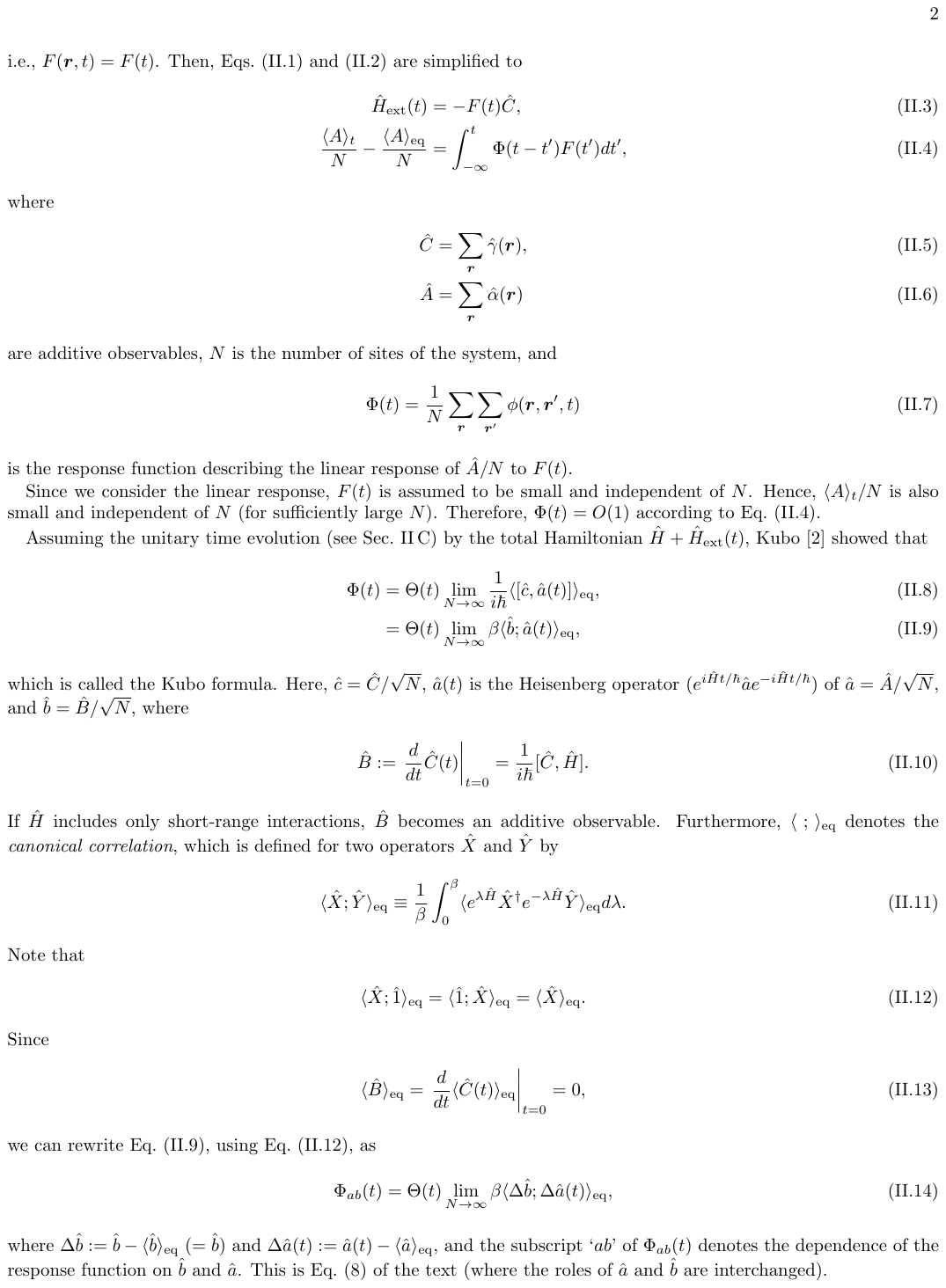}
\end{figure*}

\begin{figure*}
\vspace{-10mm}\hspace{-28mm}
\includegraphics[page=3,clip]{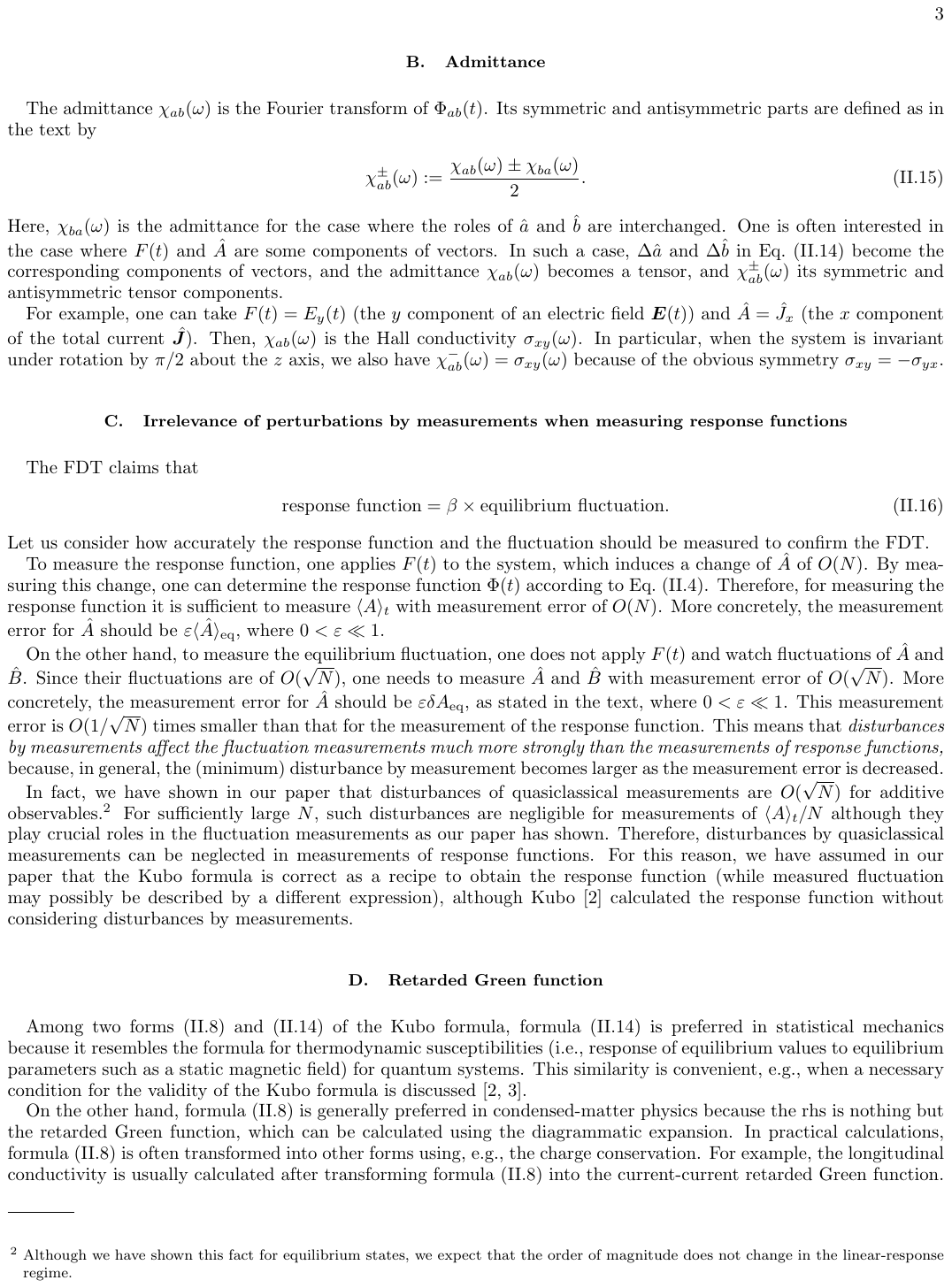}
\end{figure*}

\begin{figure*}
\vspace{-10mm}\hspace{-28mm}
\includegraphics[page=4,clip]{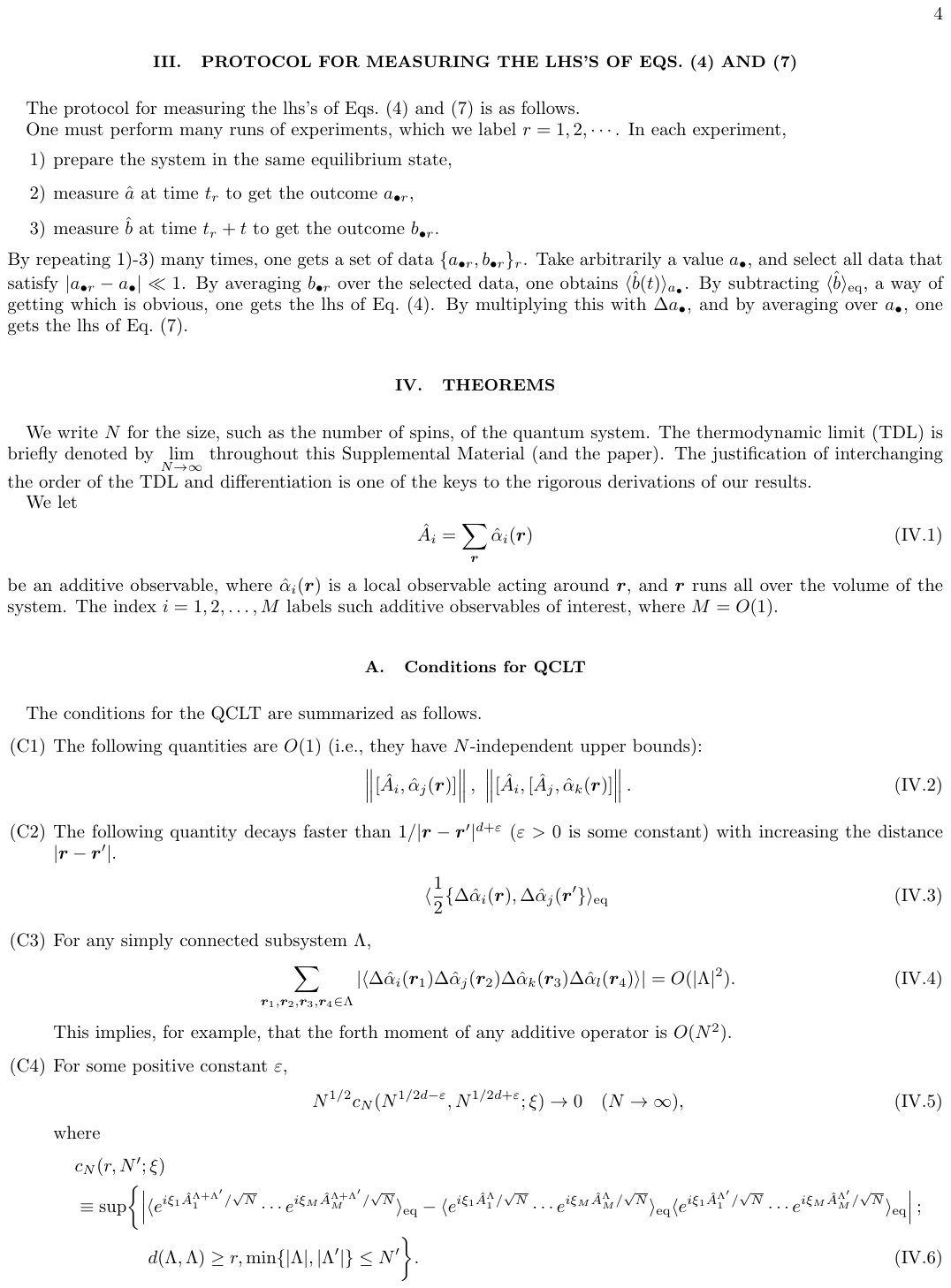}
\end{figure*}

\begin{figure*}
\vspace{-10mm}\hspace{-28mm}
\includegraphics[page=5,clip]{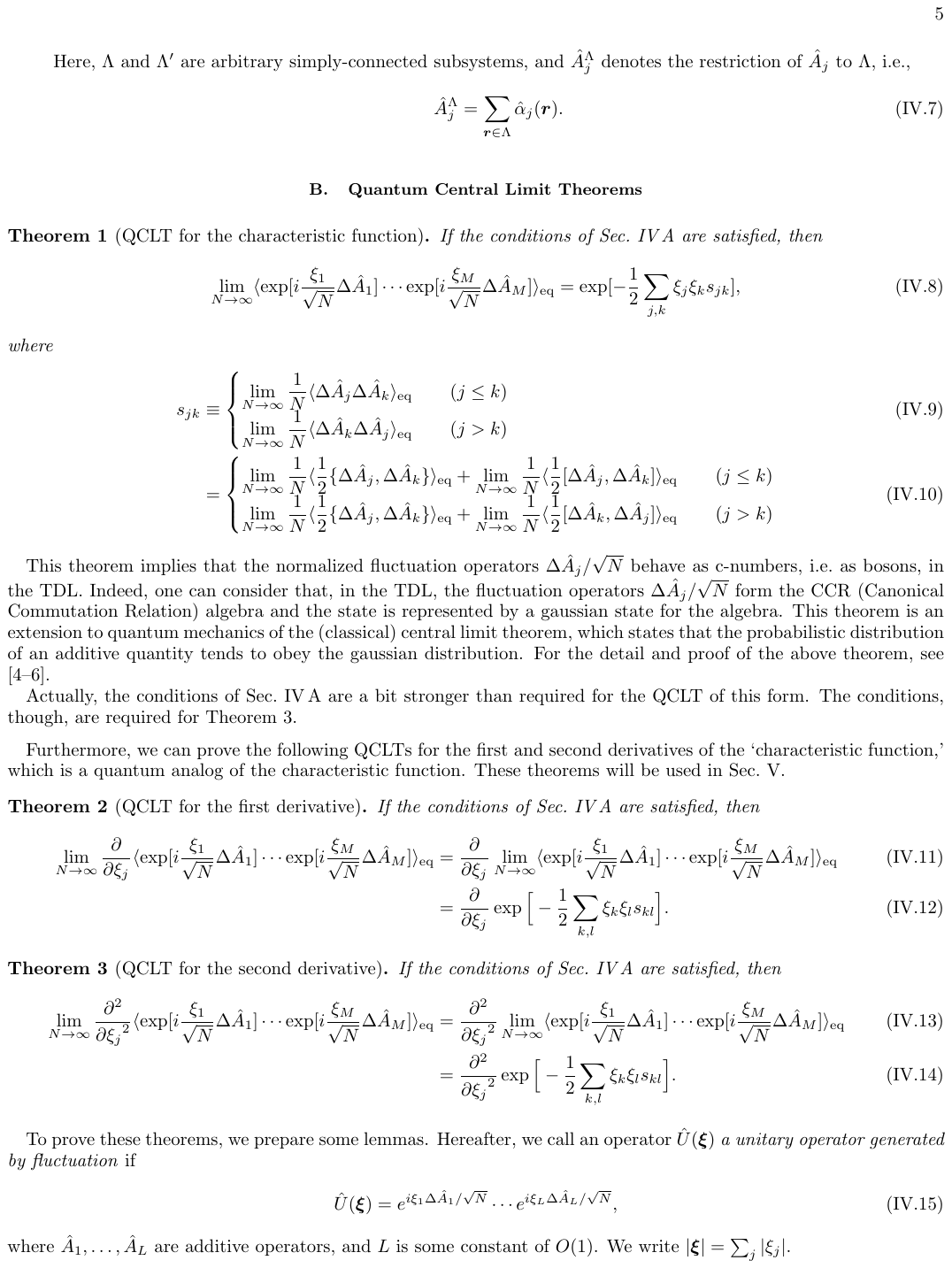}
\end{figure*}

\begin{figure*}
\vspace{-10mm}\hspace{-28mm}
\includegraphics[page=6,clip]{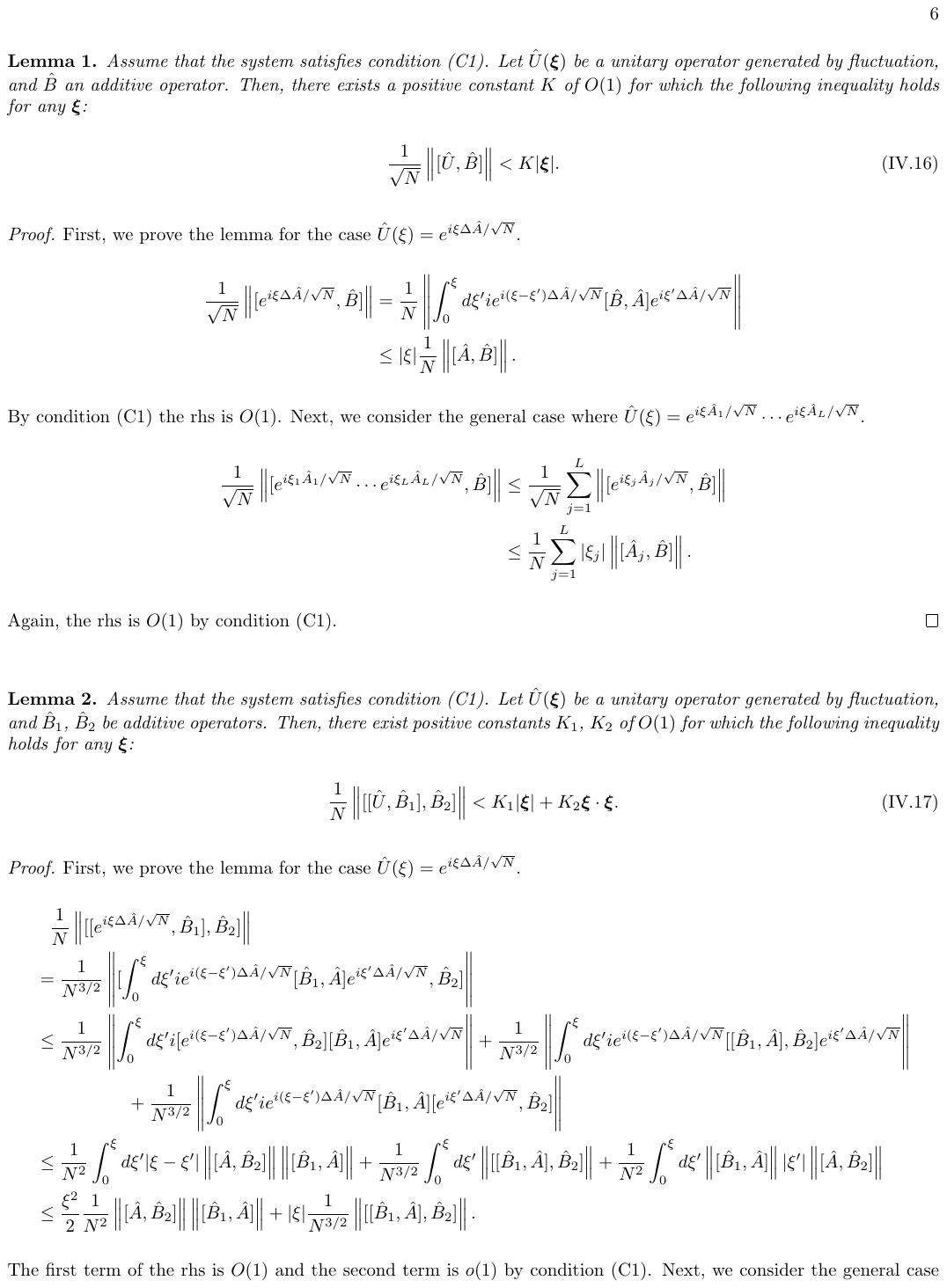}
\end{figure*}

\begin{figure*}
\vspace{-10mm}\hspace{-28mm}
\includegraphics[page=7,clip]{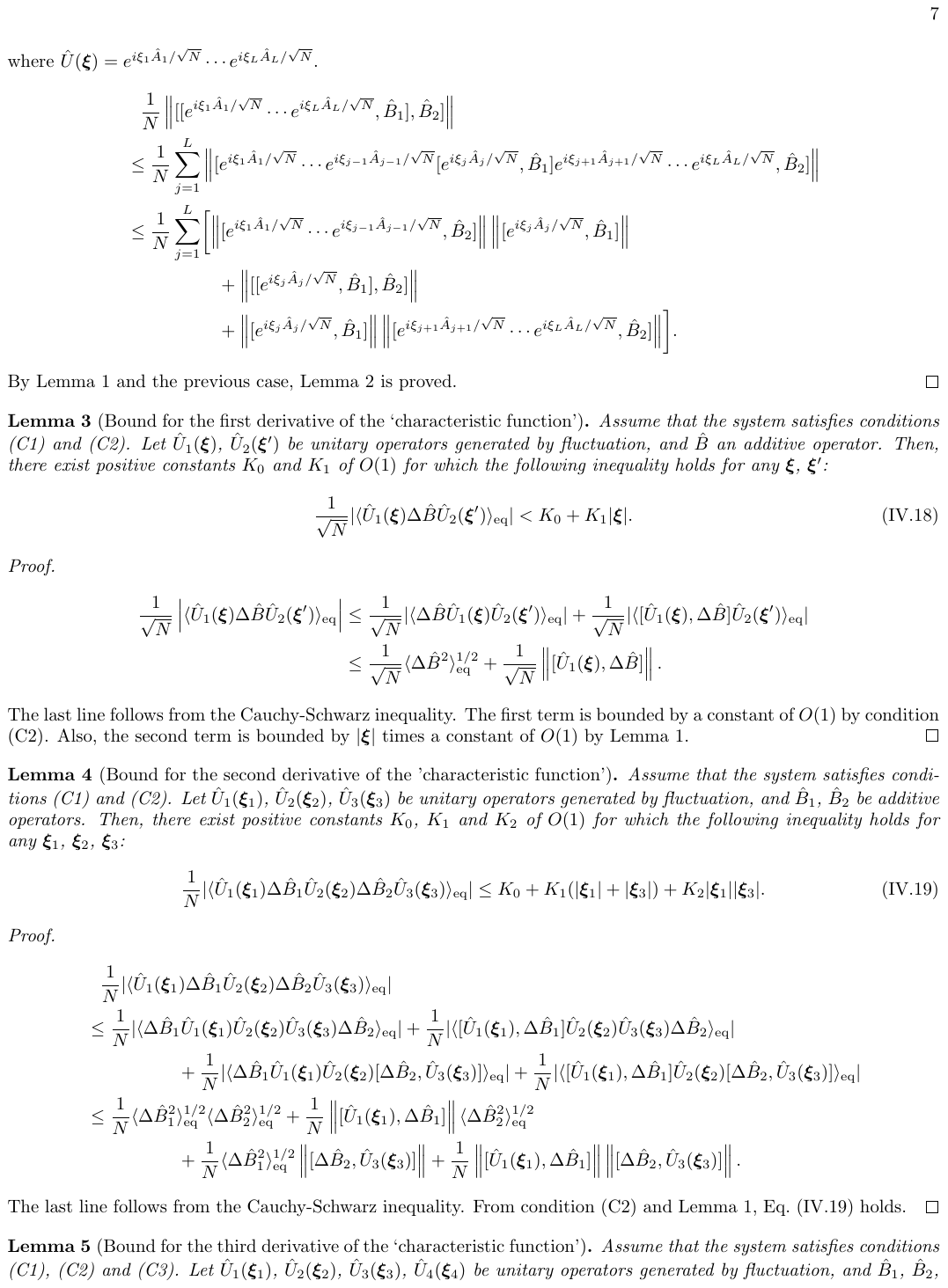}
\end{figure*}

\begin{figure*}
\vspace{-10mm}\hspace{-28mm}
\includegraphics[page=8,clip]{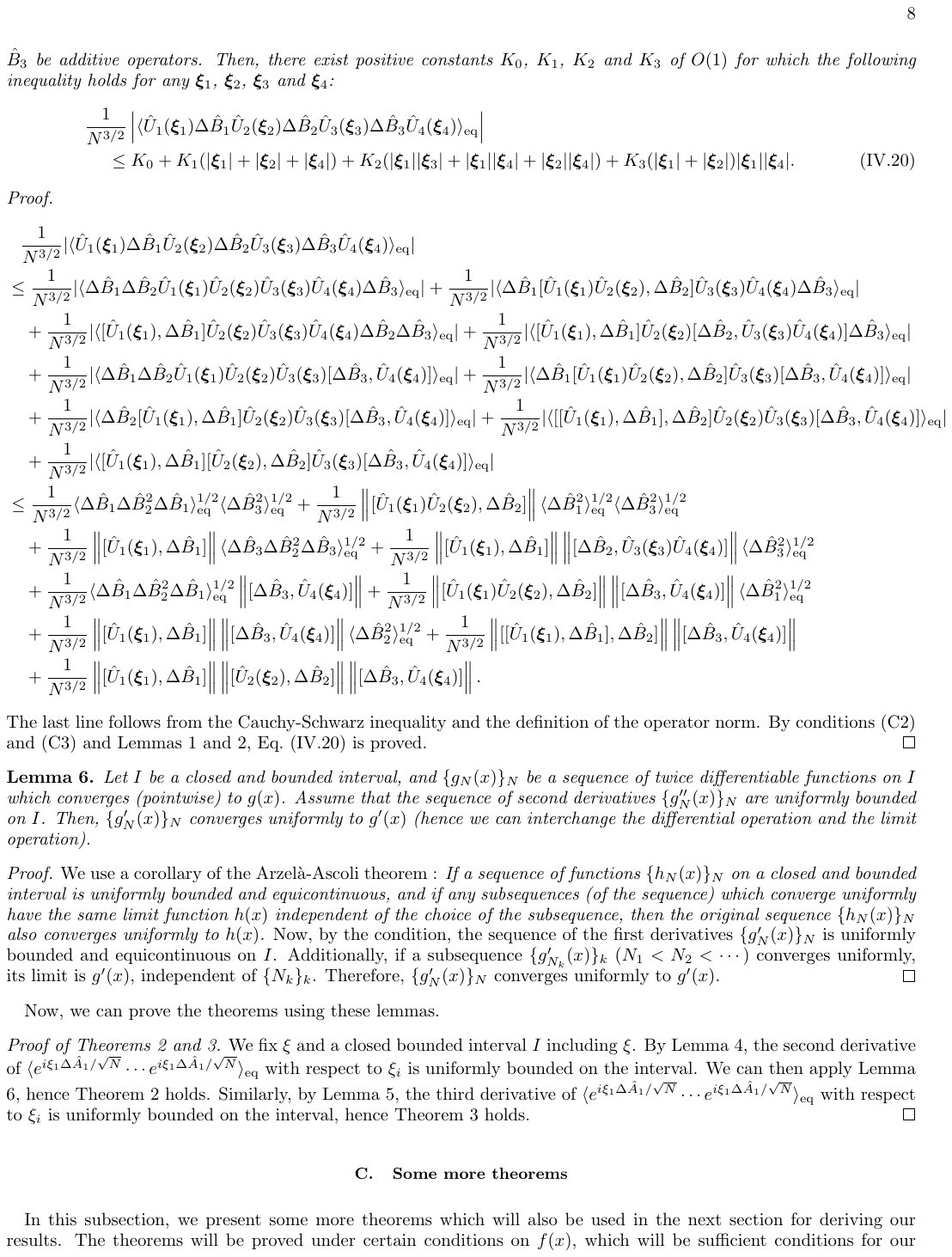}
\end{figure*}

\begin{figure*}
\vspace{-10mm}\hspace{-28mm}
\includegraphics[page=9,clip]{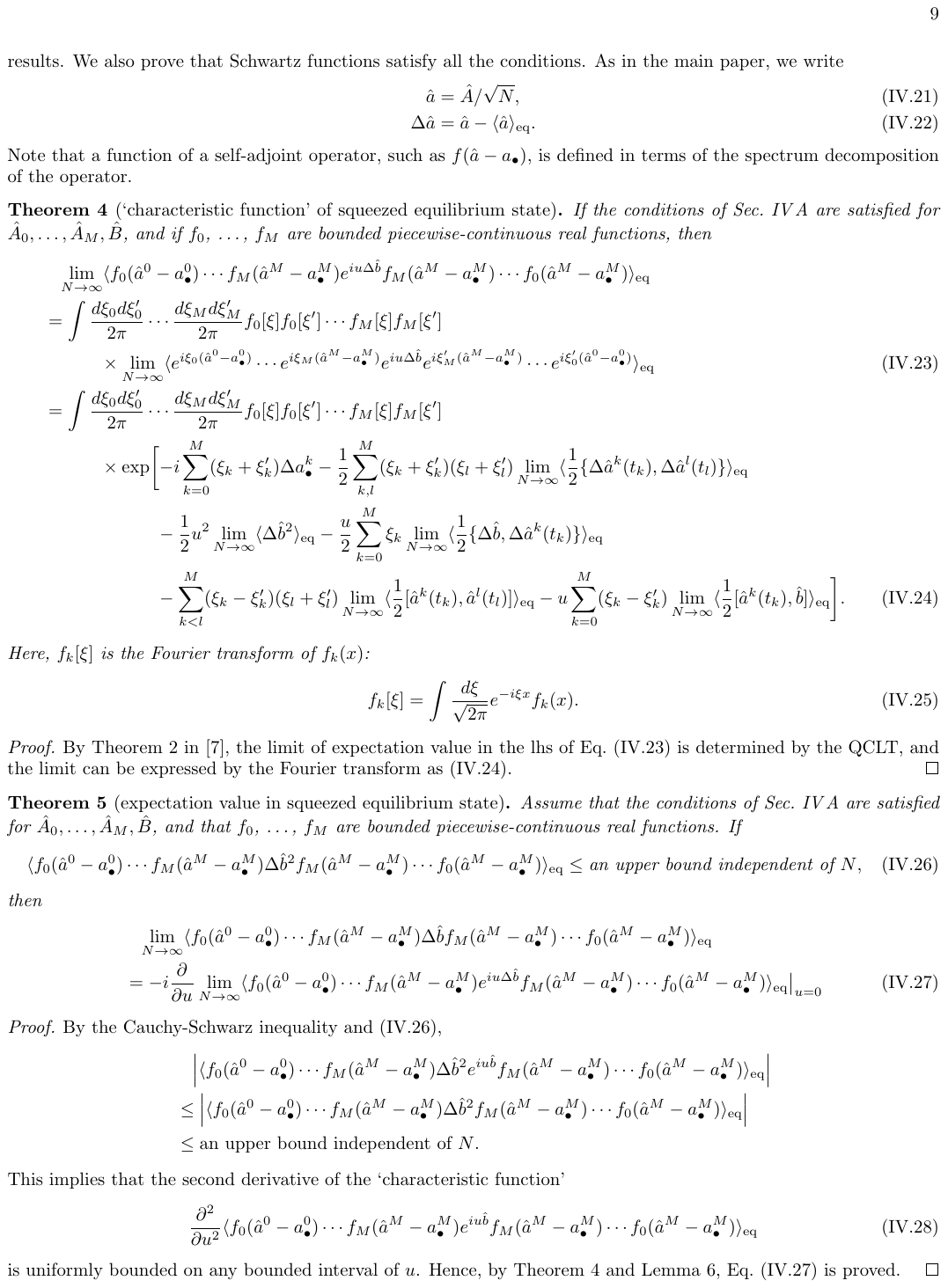}
\end{figure*}

\begin{figure*}
\vspace{-10mm}\hspace{-28mm}
\includegraphics[page=10,clip]{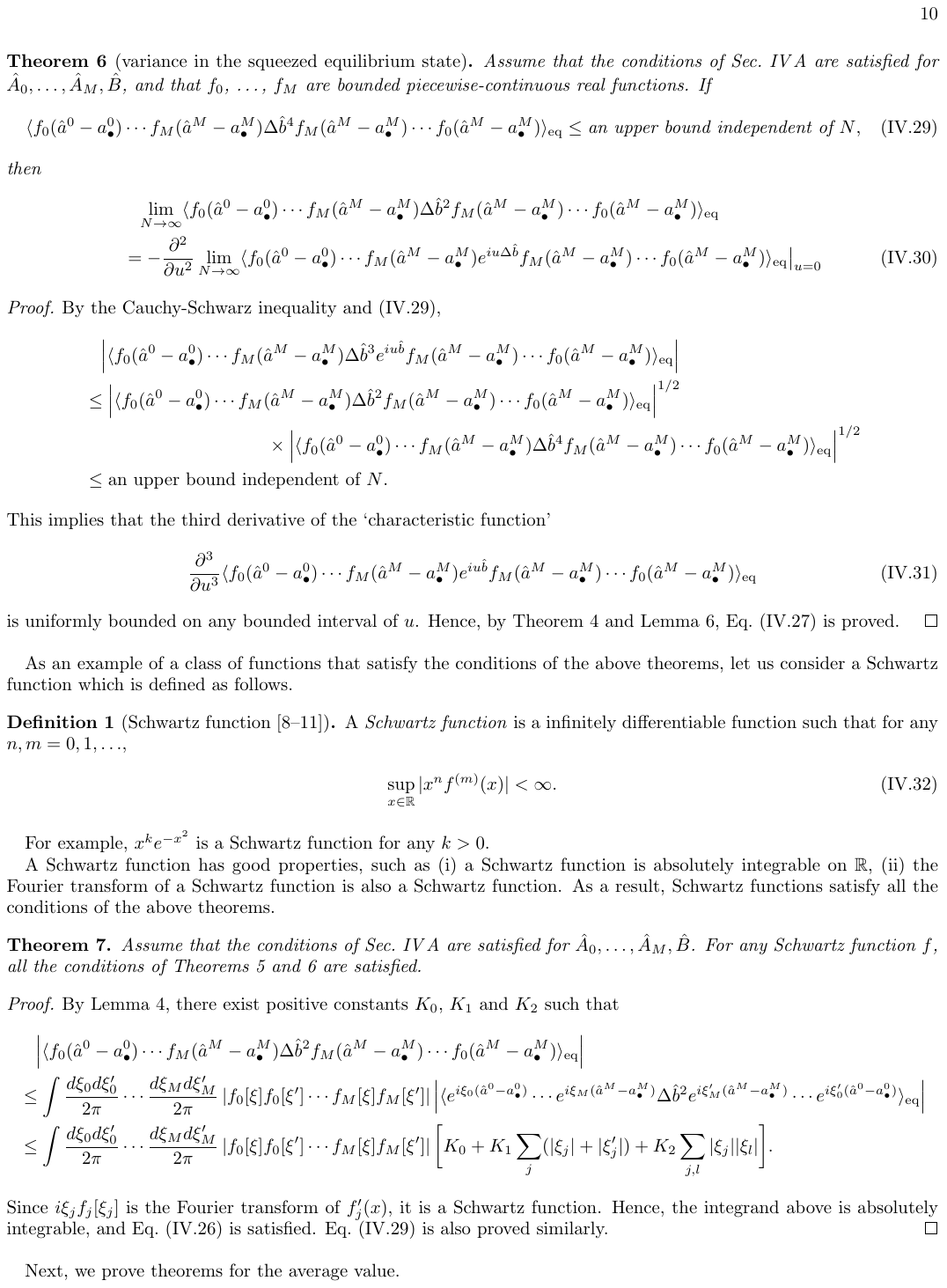}
\end{figure*}

\begin{figure*}
\vspace{-10mm}\hspace{-28mm}
\includegraphics[page=11,clip]{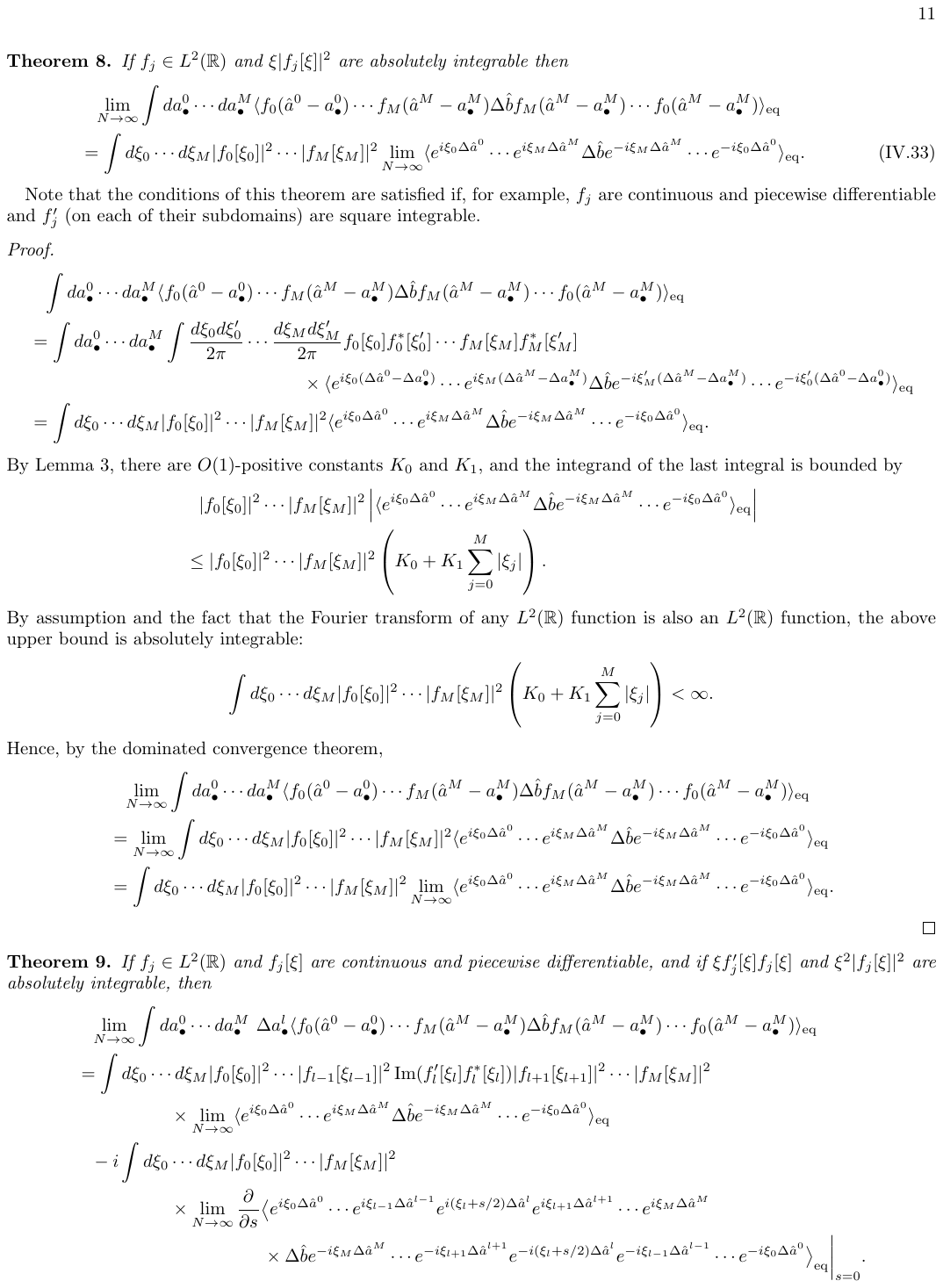}
\end{figure*}

\begin{figure*}
\vspace{-10mm}\hspace{-28mm}
\includegraphics[page=12,clip]{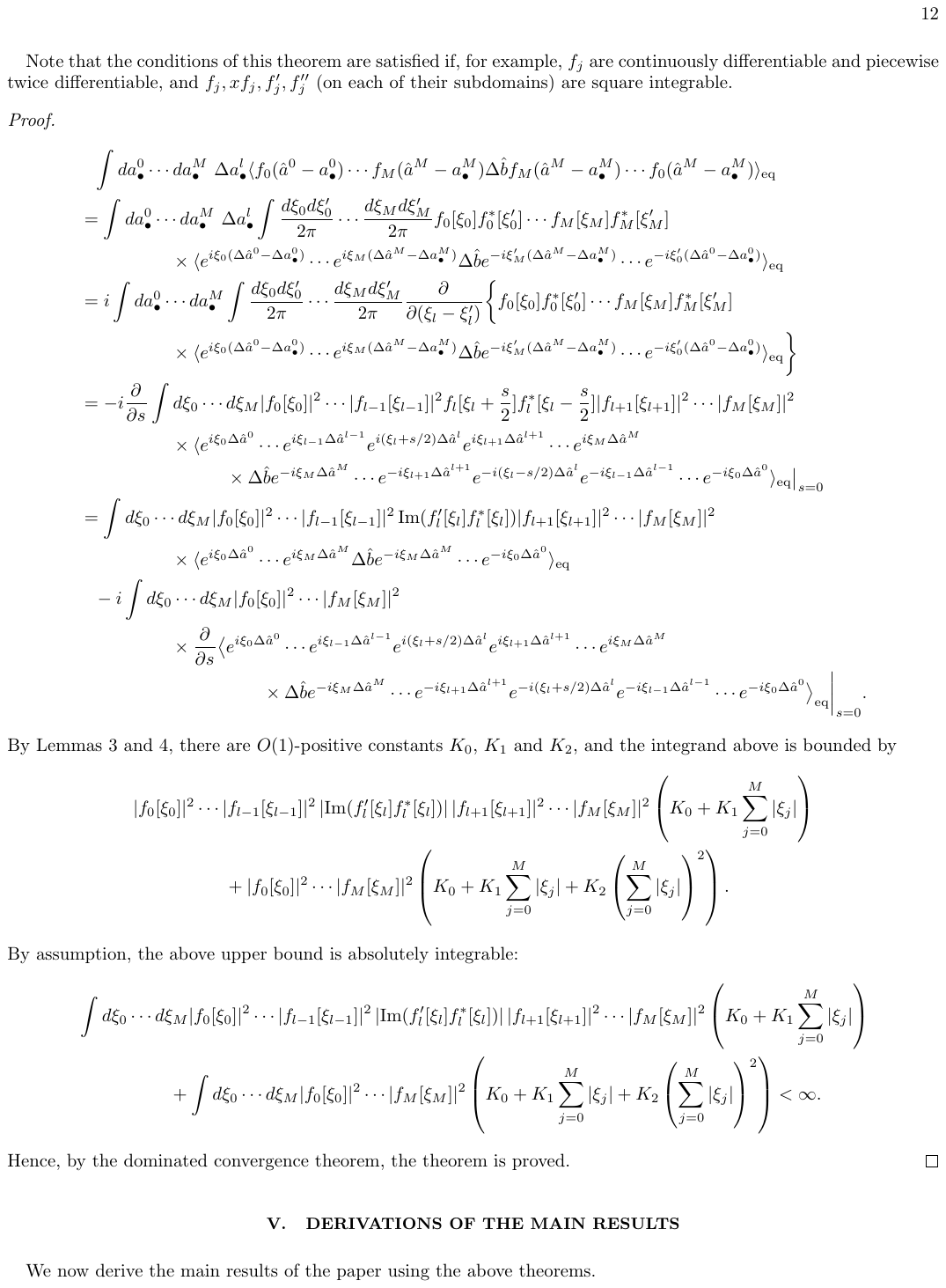}
\end{figure*}

\begin{figure*}
\vspace{-10mm}\hspace{-28mm}
\includegraphics[page=13,clip]{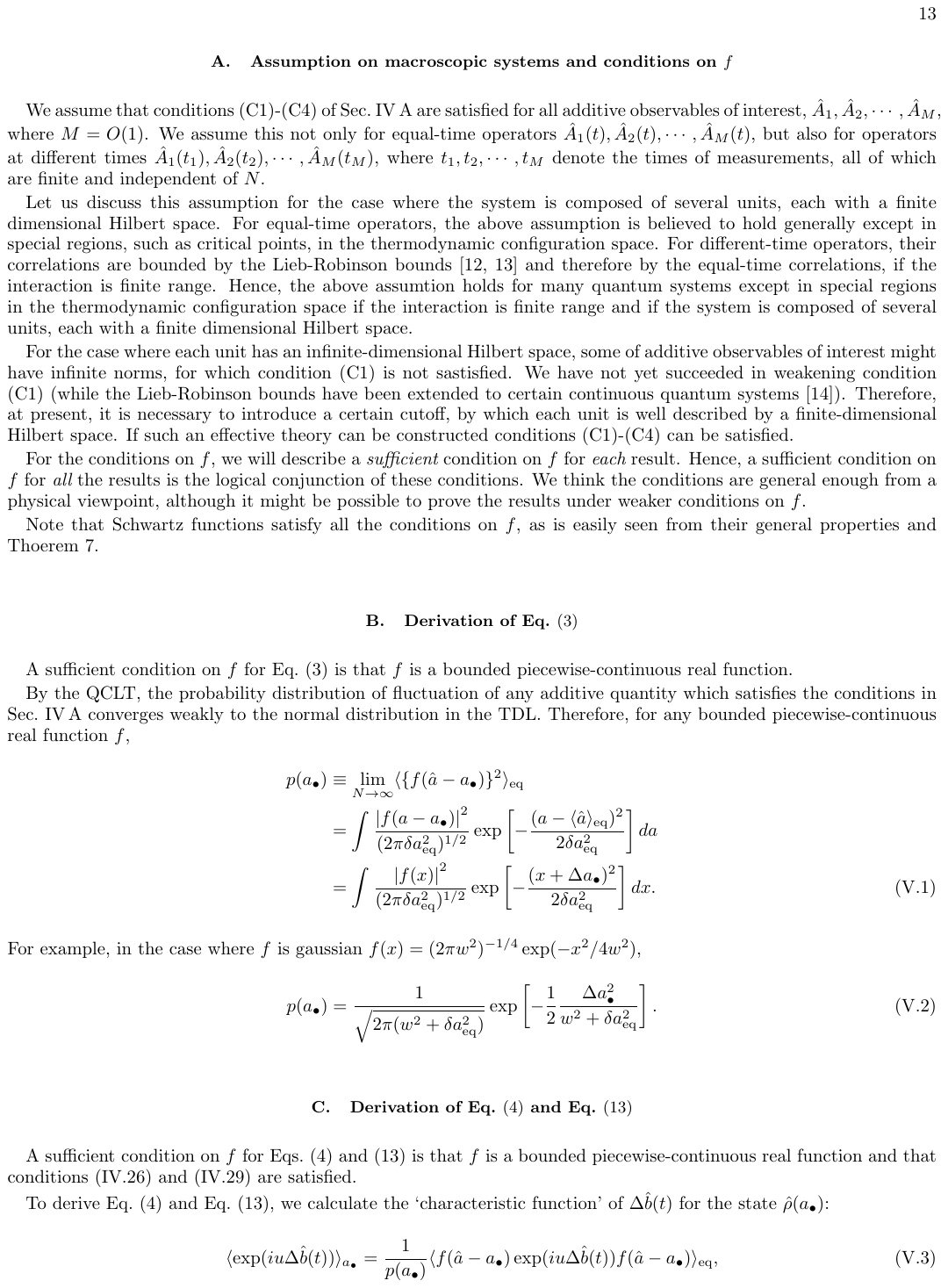}
\end{figure*}

\begin{figure*}
\vspace{-10mm}\hspace{-28mm}
\includegraphics[page=14,clip]{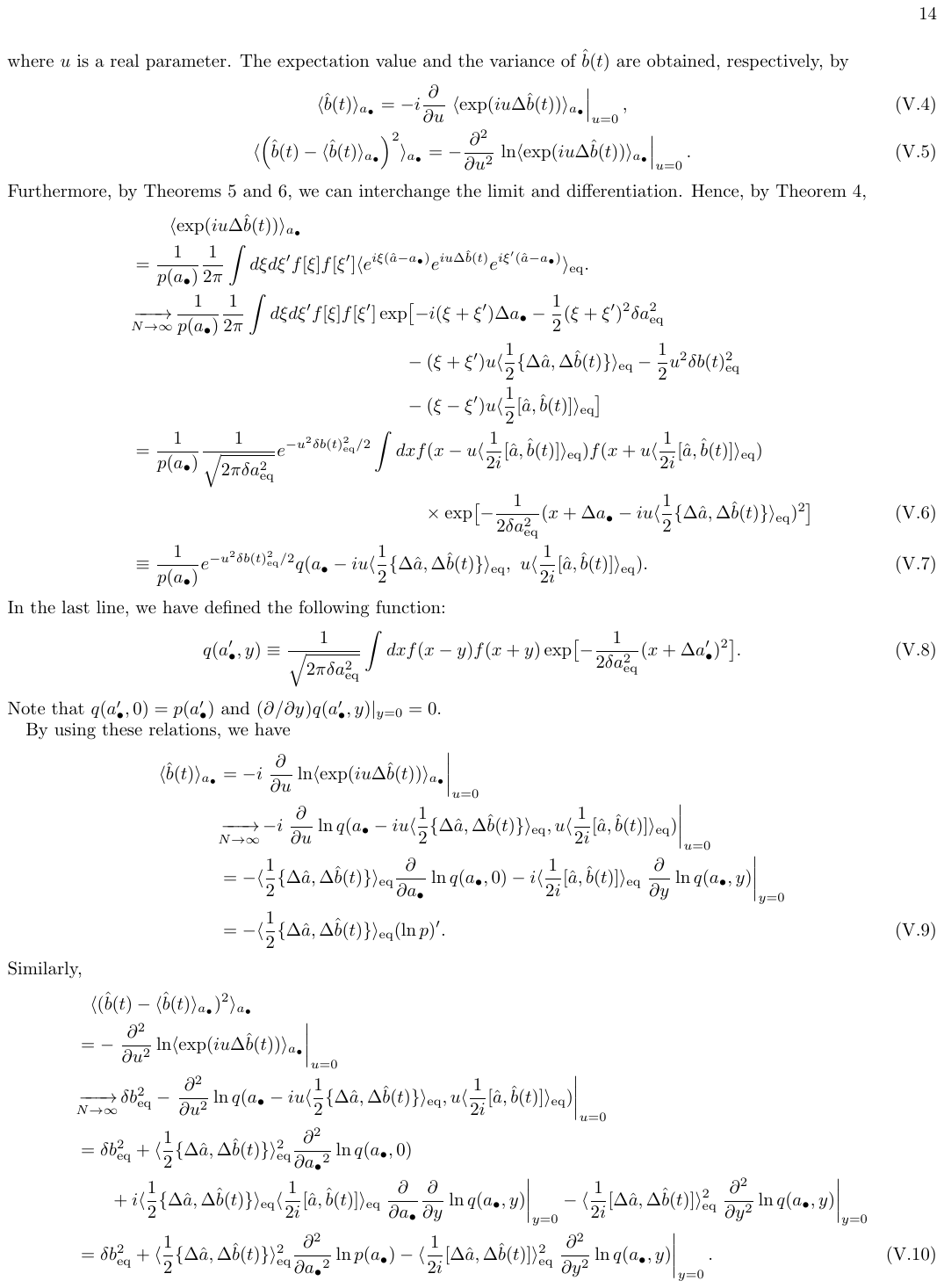}
\end{figure*}

\begin{figure*}
\vspace{-10mm}\hspace{-28mm}
\includegraphics[page=15,clip]{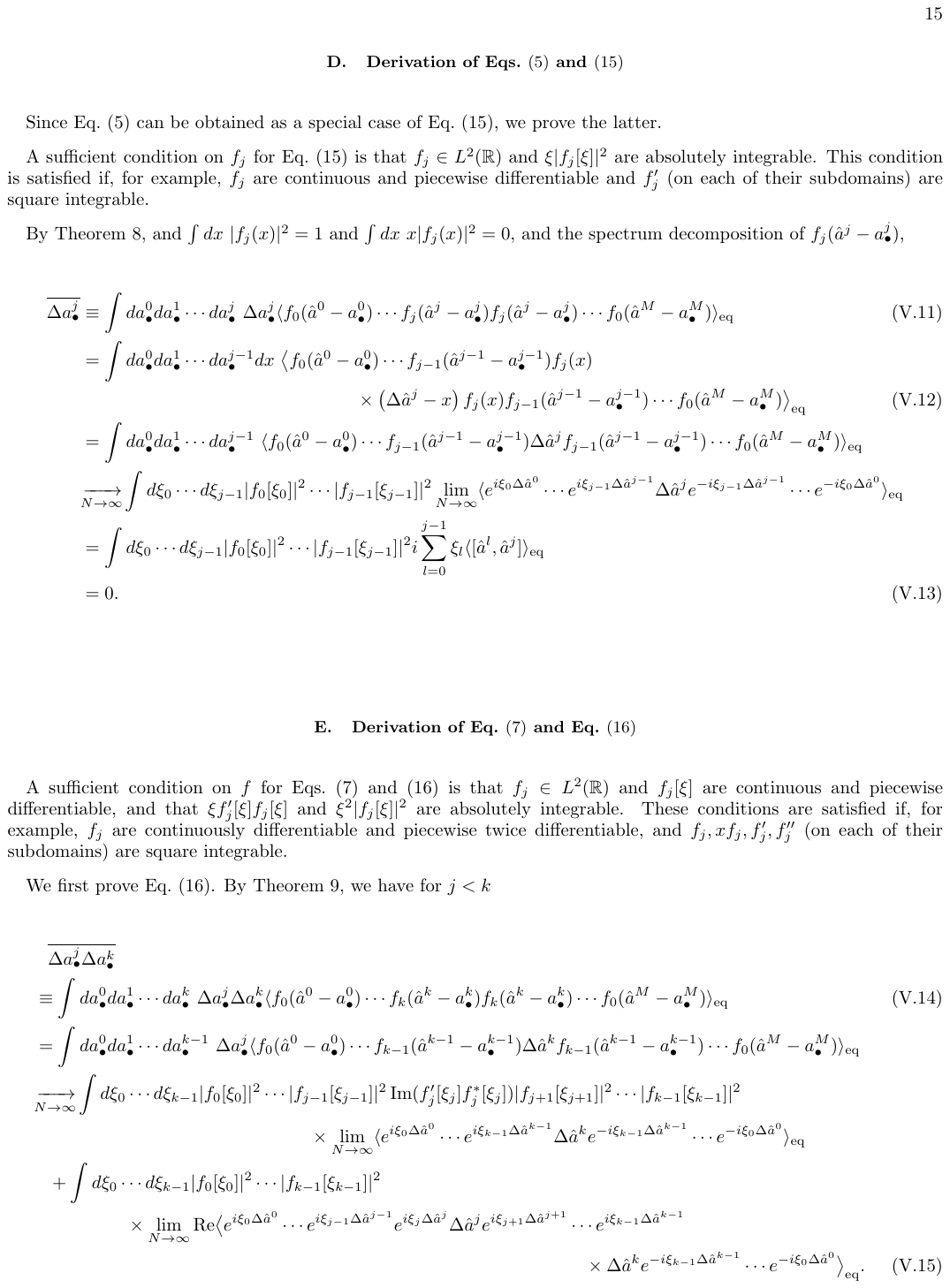}
\end{figure*}

\begin{figure*}
\vspace{-10mm}\hspace{-28mm}
\includegraphics[page=16,clip]{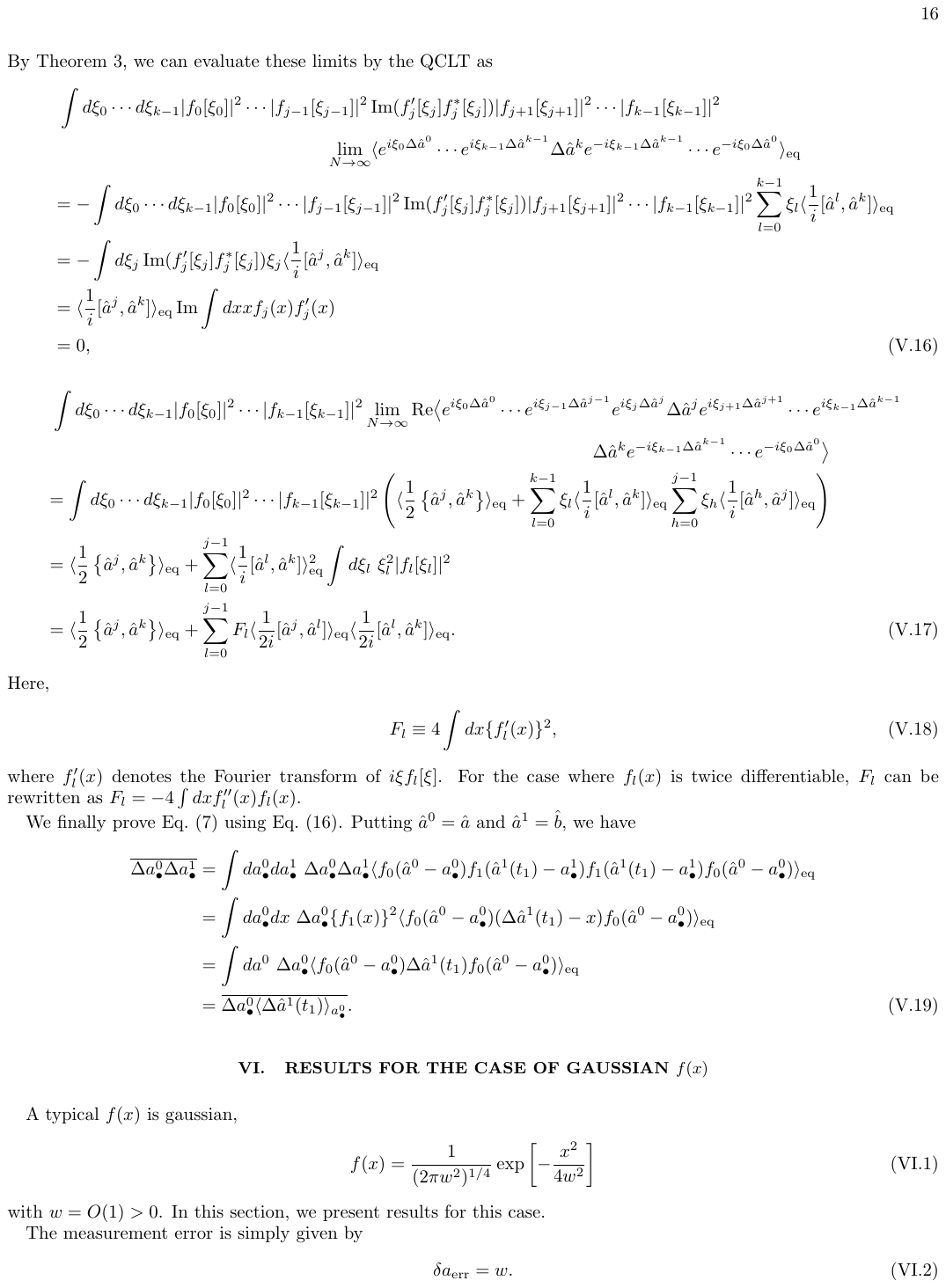}
\end{figure*}

\begin{figure*}
\vspace{-10mm}\hspace{-28mm}
\includegraphics[page=17,clip]{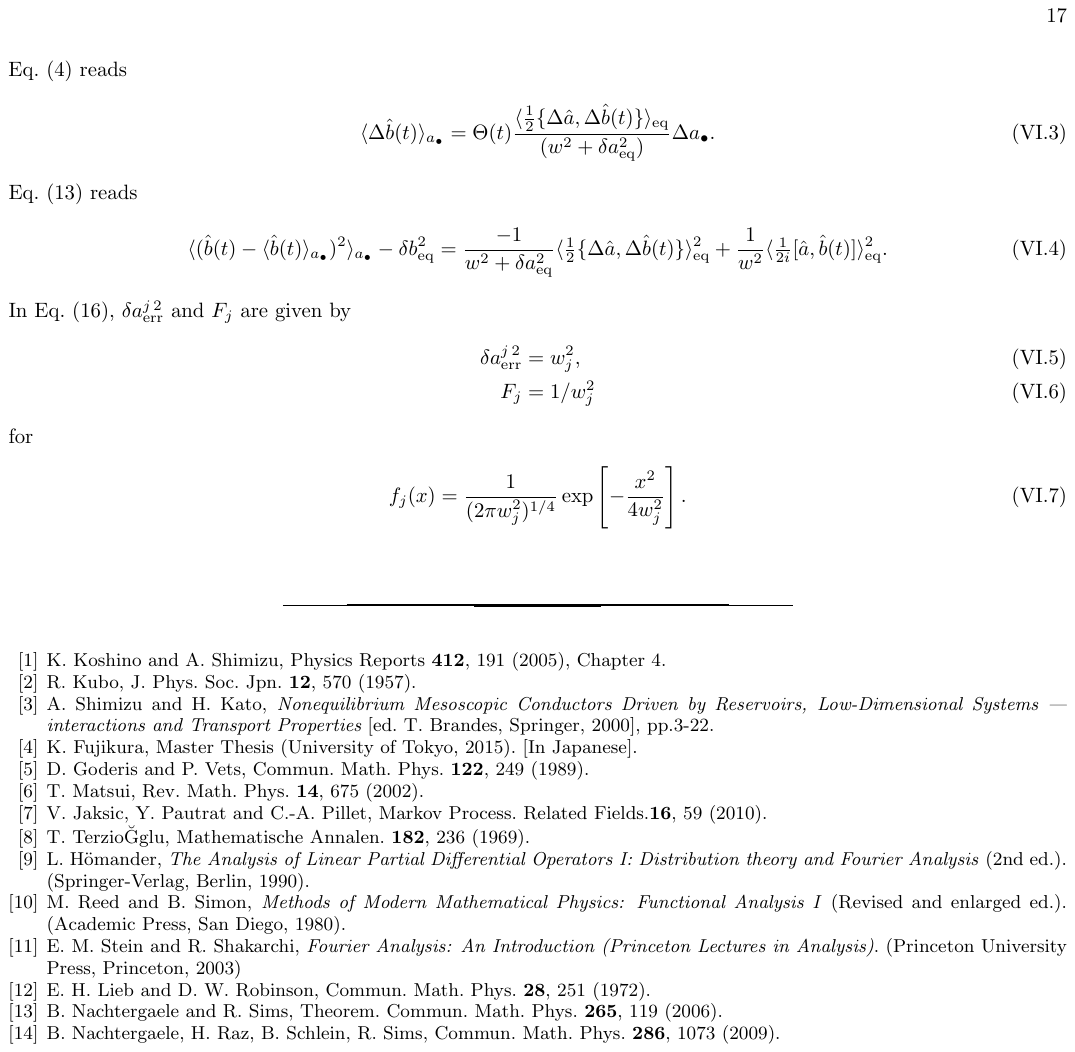}
\end{figure*}



\begin{thebibliography}{99}
\bibitem{rmk:f}
This can be generalized as in Ref.~\cite{SM}.
\bibitem{gaussian_case}
As an example, 
results for the gaussian $f$ are presented in Ref.~\cite{SM}.
\bibitem{similar}
Although similar relations between 
$\chi_{ba}$ and $\tilde{S}_{ba}$
were known \cite{KTH},
they were not taken seriously because 
{\sh physical significance of the symTC was unknown.}
\end{thebibliography}
\end{document}